\documentclass[10pt,aps,prd,twocolumn,showpacs,superscriptaddress,nofootinbib,nobibnotes,longbibliography,floatfix]{revtex4-1}

\usepackage[utf8]{inputenc}
\usepackage[T1]{fontenc}
\usepackage{bm}
\usepackage{mathtools,amsmath,amssymb,amsfonts,mathrsfs,eucal,graphicx,tensor,csquotes,accents,commath,chngcntr,siunitx}
\usepackage[dvipsnames]{xcolor}
\usepackage[unicode]{hyperref}
\hypersetup{colorlinks=true, citecolor=MidnightBlue,
            linkcolor=MidnightBlue, urlcolor=MidnightBlue, linktocpage=true}
\usepackage[normalem]{ulem}
\usepackage{orcidlink}

\newcommand{\orcid}[1]{\href{https://orcid.org/#1}{\includegraphics[width=10pt]{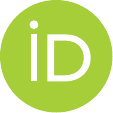}}}

\begin{document}

\author{Saulo Albuquerque \orcid{0000-0003-2911-9358}}
\affiliation{Departmento de Física, Universidade Federal da Paraíba, Caixa Postal 5008, João Pessoa 58059-900, PB, Brazil}
\affiliation{Theoretical Astrophysics, IAAT, University of Tübingen, D-72076 Tübingen, Germany}
\email{saulosoaresfisica@gmail.com}

\author{Sebastian H. V\"olkel \orcid{0000-0002-9432-7690}}
\affiliation{Max Planck Institute for Gravitational Physics (Albert Einstein Institute), D-14476 Potsdam, Germany}
\email{sebastian.voelkel@aei.mpg.de}

\author{Kostas D. Kokkotas \orcid{0000-0001-6048-2919}}
\affiliation{Theoretical Astrophysics, IAAT, University of Tübingen, D-72076 Tübingen, Germany}

\author{Valdir B. Bezerra \orcid{0000-0001-7893-0265}}
\affiliation{Departmento de Física, Universidade Federal da Paraíba, Caixa Postal 5008, João Pessoa 58059-900, PB, Brazil}

\date{\today}

\title{Inverse problem of analog gravity systems}

\begin{abstract}
Analog gravity models of black holes and exotic compact objects provide a unique opportunity to study key properties of such systems in controlled laboratory environments. 
In contrast to astrophysical systems, analog gravity systems can be prepared carefully and their dynamical aspects thus investigated in unprecedented ways. 
While gravitational wave scattering properties of astrophysical compact objects are more connected to quasinormal modes, laboratory experiments can also access the transmission and reflection coefficients, which are otherwise mostly relevant for Hawking radiation related phenomena. 
In this work, we report two distinct results. 
First, we outline a semiclassical, nonparametric method that allows for the reconstruction of the effective perturbation potential from the knowledge of transmission and reflection coefficients for certain types of potentials in the Schr\"odinger wave equation admitting resonant tunneling. 
Second, we show how to use our method by applying it to an imperfect draining vortex, which has been suggested as an analog of extreme compact objects. 
Although the inverse problem is ,in general, not unique, choosing physically motivated assumptions and requiring the validity of semiclassical theory, we demonstrate that the method provides efficient and accurate results. 
\end{abstract}

\maketitle

\section{Introduction}\label{intro}

Since the direct measurements of gravitational waves from binary mergers of black holes and neutron stars, it is finally possible to explore strong field dynamics in a direct way~\cite{LIGOScientific:2016aoc,LIGOScientific:2017ycc,LIGOScientific:2018mvr,LIGOScientific:2020ibl,LIGOScientific:2021djp}. 
Future improvements of existing detectors and promising next generation successors will provide us with pristine tests of general relativity and explore compact objects~\cite{LISA:2017pwj,Maggiore:2019uih,Reitze:2019iox}. 
As such measurements originate from astrophysical sources, initial conditions and properties of the systems cannot be explored in an arbitrary way, but are ultimately given by whatever is realized in nature. 
In particular from a theoretical point of view, certain types of observables cannot be directly probed in this context, but are of fundamental interest. 

Within compact object perturbation theory, it is known that the dynamics of perturbed fields or the metric can often be cast in the form of one-dimensional wave equations with a potential term similar to those studied in quantum mechanics~\cite{Regge:1957td,Zerilli:1970se,Teukolsky:1973ha}. 
In addition to the question of the eigenvalue spectrum of a given potential, one of the most common problems is also computing transmission and reflection coefficients. 
The former one manifests itself in the calculation of quasinormal modes~\cite{Kokkotas:1999bd,Berti:2009kk,Konoplya:2011qq,Franchini:2023eda}, which are relevant for the ringdown phase of a binary merger and of special interest for testing the assumptions of the Kerr hypothesis~\cite{Carter:1971zc,Robinson:1975bv,Kerr:1963ud}. 
In astrophysical scenarios, transmission and reflection coefficients cannot directly be extracted. In fact, they are more accurately described by a controlled comparison of ingoing and outgoing radiation. 
It may be indirectly addressable with extreme mass ratio inspirals (EMRIs) with future detectors like the Laser Interferometer Space Antenna~\cite{LISA:2017pwj}, in which the smaller object can be treated perturbatively around a massive exotic compact object, e.g., see Refs.~\cite{Maggio:2021uge,Cardoso:2022fbq}. 
Moreover, very recently it was demonstrated that it may also be possible to extract graybody factors from ringdown of EMRIs, at least approximately~\cite{Oshita:2023cjz}.  

Analog gravity provides exciting and complementary ways to study qualitatively similar phenomena and systems ~\cite{visser1,visser2,cardosoanalogue}, but based on much simpler underlying physics. For an extensive review, see Ref.~\cite{Barcelo:2005fc}. 
One well-known example is Hawking radiation, which cannot be directly measured for astrophysical black holes, but can be mimicked in analog systems~\cite{Unruh:1980cg,hawkingbec1,hawkingbec2,hawkingbec3}, as has been suggested and studied in Bose-Einstein condensates, both theoretically~\cite{bec1,bec2,kostashoracio}, and experimentally~\cite{Steinhauer:2014dra,Steinhauer:2015saa}. 
Backreaction effects in a hydrodynamical setup have been studied in Ref.~\cite{Patrick:2019kis}. 
Another black hole related observable are quasinormal modes, which have been investigated experimentally in an analog black hole setup in Ref.~\cite{Torres:2020tzs}. 

In recent years, the study of astrophysical exotic compact objects has received much attention, as some works claimed tentative evidence of smoking-gun signs of so-called echoes in gravitational wave data, e.g., Ref.~\cite{Abedi:2016hgu}. 
The echo phenomenon was first studied for ultracompact constant density stars in Refs.~\cite{doi:10.1098/rspa.1991.0117, Kokkotas:1995av,Tominaga:1999iy,Ferrari:2000sr}, and, since then, for a variety of exotic compact objects. See Ref.~\cite{Cardoso:2019rvt} for a review on the topic. 
Although subsequent works cannot confirm such findings and refuse claimed significance~\cite{LIGOScientific:2020tif,LIGOScientific:2021sio}, the question of the existence of such objects remains intriguing. 
In the context of analog gravity, a system with such properties has been proposed in Ref.~\cite{Torres:2022bto} and consists of an imperfect draining vortex. 
In the same work, the observational consequences of such a system have been studied for scalar field perturbations. 
These include transmission and reflection coefficients, as well as possible superradiant features, whose formal treatment, we call the \textit{direct problem} in the following. 
If analog exotic compact objects can be realized in laboratory experiments and transmission/reflection coefficients can be measured, how could one use them to study their properties in the \textit{inverse problem}?

In this work, we present a novel method that is based on a semianalytic analysis of the underlying wave equation and that does not require the specification of the details of the underlying model. 
Note that this is very different from standard inference approaches, in which one reconstructs the parameters of a model using statistical tools. 
Instead, we show how Wentzel-Kramers-Brillouin (WKB) theory-based results known for quasistationary states of astrophysical exotic compact objects~\cite{Volkel:2017kfj,Volkel:2018hwb,Volkel:2019ahb} can be extended to recover main properties of the effective potential, as well as absorption properties at the surface of the objects. 
As proof of principle, we apply it to the imperfect draining vortex model studied in Ref.~\cite{Torres:2022bto}, for which we compute the relevant observables with standard, accurate numerical methods. 
Since inverse problems are often not uniquely solvable, we argue how physically motivated constraints allow one to reconstruct the relevant potential. 
The quality of the reconstruction, due to its relation to WKB theory, is very good for large angular numbers. 
We are also able to reconstruct the reflectivity of the objects, for which results become more accurate the more the object is reflecting incoming waves. 

This work is structured as follows. 
In Sec.~\ref{methods}, we first outline the methods to solve the inverse problem for given transmission and reflection functions. 
We then apply these methods to the imperfect draining vortex model and discuss our results in Sec.~\ref{app_results}. 
Finally, our conclusions can be found in Sec.~\ref{conclusions}.

\section{Methods}\label{methods}

In this section we first outline the fundamentals of the direct problem in Sec.~\ref{direct_problem}, then we present the inversion of the WKB-based methods in Sec.~\ref{BS_Gamow} and, finally, discuss in Sec.~\ref{analysis_transmission} how the transmission and reflection functions need to be further analyzed in order to provide the input for the analysis presented in the former section.

\subsection{Outline of the direct problem}\label{direct_problem}

Throughout this work, the main properties of the systems we consider can be obtained by studying the following effective one-dimensional wave-equation:
\begin{align}\label{wave_eq}
\frac{\text{d}^2}{\text{d}x^2}\psi(x) + \left[E-V(x) \right]\psi(x)=0.
\end{align}
Here $V(x)$ is, in general, an energy-dependent potential that captures the dynamical properties of the object under consideration. 
Exotic astrophysical systems~\cite{Cardoso:2019rvt}, as well as the analog systems we study in Sec.~\ref{app_results} can be best described by a potential barrier with model-dependent reflection properties on one side of the barrier. 
To understand the description of the inverse problem, let us first review the key concepts of the direct one. 

There are two common scenarios in which Eq.~\eqref{wave_eq} is typically studied. 
One of them is an eigenvalue problem for discrete values of $E_n$ that are determined from suitably chosen boundary conditions. 
In its most basic form, this can either give bound states (purely real eigenvalues of potential wells) or quasinormal modes (complex eigenvalues of potential barriers). 
The second scenario is the scattering problem of transmission and reflection coefficients, which is more commonly studied in quantum mechanics or in the context of black holes for Hawking radiation calculations. 
Both scenarios are not independent of each other, and in fact, our framework requires a joint analysis to address the inverse problem.

\subsubsection{Semiclassical method}

\begin{figure}
\includegraphics[width=1.0\linewidth]{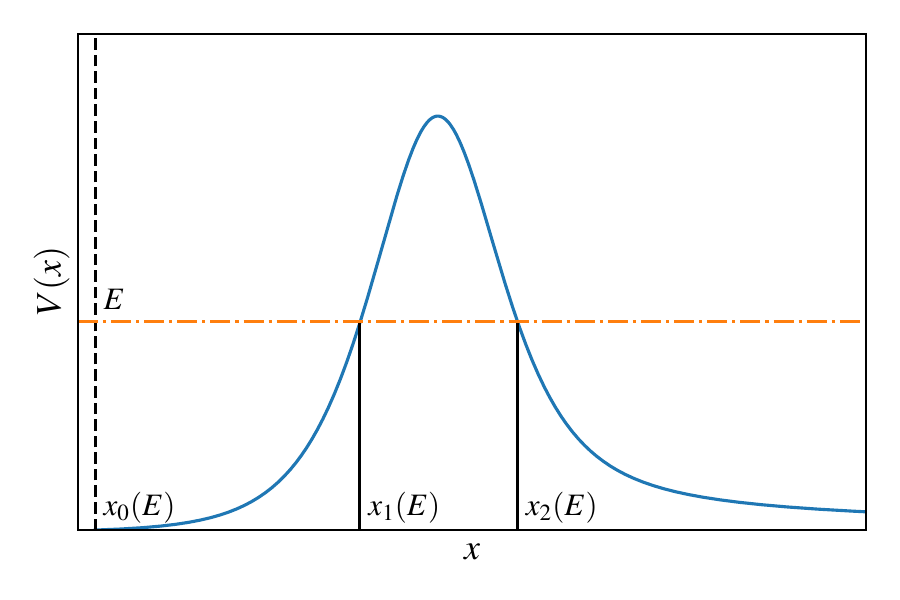}
\caption{Here we show a typical potential barrier $V(x)$ (blue) with turning points $x_0(E), x_1(E)$, and $x_2(E)$ for a given value of $E$ (orange dot-dashed line). The location of the reflective core's surface coincides with $x_0(E)$ (black dotted line).}
\label{fig_potential}
\end{figure}

For those astrophysical or analog systems for which the outlined method here is valid, the typical structure of the potential yields the so-called quasistationary states as eigenvalue problem. 
The spectrum $\omega_n^2 = E_n = E_{0n} + \mathrm{i} E_{1n}$ of those modes is characterized by real valued bound states $E_{0n}$, together with a very small imaginary part $E_{1n}$ reflecting the transmission through the barrier and ``surface'' thus measuring the respective mode's lifetime. 
In the astrophysical context, these modes have first been found for ultracompact constant density stars in Refs.~\cite{doi:10.1098/rspa.1991.0104,Kokkotas:1994an} and are also known as trapped modes. 
Accordingly, as we usually have in the context of compact object perturbation theory, the real part of $\omega_n$ describes the frequency of the $n$ th mode, while the imaginary part of $\omega_n$ is inversely proportional to the damping time of that respective mode. 
Therefore, exponentially small imaginary parts imply long-living trapped modes. 
This is physically expected in a potential well created between a reflective surface and a potential barrier. 
We show such a typical case in Fig.~\ref{fig_potential}.

As we will see later in the results, increasing the reflectivity of the reflective wall tends to increase the lifetime of the trapped modes. 
Since a ``larger portion'' of the wave is being reflected by the compact object's surface, rather than being absorbed, those waves will be trapped in the well for a longer time before they actually manage to escape the well (being absorbed by the object, or being sent back to infinity). Therefore, larger reflectivity in the surface of the compact object implies exponentially smaller imaginary parts for the modes, which in turn leads to narrower widths in the transmission plots, as we discuss and illustrate later.

A semianalytic treatment of astrophysical exotic compact objects with such properties has been studied in Ref.~\cite{Volkel:2017ofl}, which combined the classical Bohr-Sommerfeld rule 
\begin{align}\label{cBS}
\int_{x_0}^{x_1} \sqrt{E_{0n}-V(x)} \text{d}x = \pi \left(n+\frac{1}{2} \right),
\end{align}
and the Gamow formula
\begin{align}
E_{1n} =& - \frac{1}{2}
\left(T_1(E) + T_2(E)\right) \left(\int_{x_0}^{x_1}\frac{1}{\sqrt{E_n-V(x)}} \text{d}x \right)^{-1}.\label{E1n_1}
\end{align}
Here $x_0, x_1, x_2$ are the classical turning points defined by $E_{0n} = V(x)$. 
The semiclassical approximations for the transmissions $T_2(E)$ through a potential barrier is given by
\begin{align}
T_2(E) &= \exp\left(2 i \int_{x_{1}}^{x_{2}} \sqrt{E_n-V(x)} \text{d}x\right),
\label{transmission}
\end{align}
while $T_1(E)$ is defined in terms of boundary conditions at the object's surface located at $x_0$ and discussed in Sec.~\ref{vortex}. 
See also Ref.~\cite{Cardoso:2014sna} for a very similar approach for the direct problem. 
The inverse problem related to reconstructing properties of $V(x)$ given the spectrum of quasistationary states was studied in Refs.~\cite{Volkel:2017kfj,Volkel:2018hwb} by inverting the Bohr-Sommerfeld rule and Gamow's formula, as is explained in more detail in Sec.~\ref{BS_Gamow}. 

\subsubsection{Numerical method}\label{methods_numerical}

Since the main focus of this work is to study the inverse problem from the transmission and reflection coefficients, we conclude this section with a summary of the approach to solve the direct problem numerically. 
We refer the interested reader to Ref.~\cite{Torres:2022bto} for more details. 
The results of the direct study of the scattering problem then provide us with the starting point for our study of the inverse problem. 

The boundary condition that needs to be incorporated at the inner boundary is given by
\begin{align}\label{BCs1}
\psi(x \approx x_0) \sim  A^\text{wall} \left[e^{- \mathrm{i} \tilde{\omega}x} 
+ K e^{- 2 \mathrm{i} \tilde{\omega}x_0 }  e^{ \mathrm{i} \tilde{\omega}x} \right],
\end{align}
with $\tilde{\omega} = \omega - m C$, and where $K$ is the reflectivity constant at the wall in $x_{0}$. 
At spatial infinity, they are given by
\begin{align}\label{BCs2}
\psi(x \rightarrow \infty) \sim A^\text{in} e^{-\mathrm{i} \omega x} +A^\text{out} e^{+\mathrm{i} \omega x}.
\end{align}
From the amplitudes $(A^\text{in},A^\text{out})$, we can define the reflection and transmission coefficients by
\begin{align}
|t|^2&= \frac{|A^\text{wall}|^2}{|A^\text{in}|^2}(1-|K|^2), \\  |r|^2&= \frac{|A^\text{out}|^2}{|A^\text{in}|^2}.
\end{align}
They are related to each other by
\begin{align}
|r|^2= 1-\frac{\tilde{\omega}}{\omega}|t|^2.
\end{align}
Note the notation $T(E) = t(E)^2$ as used in Eqs.~\eqref{E1n_1} and \eqref{transmission}. 
As usual, we know that the physical solutions satisfying the boundary conditions given by eq.~\eqref{BCs1}  
and eq.~\eqref{BCs2} can be decomposed into the basis of solutions $(u_{h},u_{\infty})$, defined by the following asymptotic behavior:
\begin{align}
u_{h}\sim \biggl\{\begin{array}{ll} e^{- \mathrm{i} \tilde{\omega}x},&   x\rightarrow - \infty, 
\\
A^{-}_{\infty}e^{- \mathrm{i} \omega x}+ A^{+}_{\infty}e^{ \mathrm{i} \omega x}, &      
 x\rightarrow + \infty,\end{array}
\end{align}
and
\begin{align}
u_{\infty}\sim \biggl\{\begin{array}{ll} A^{-}_{h}e^{- \mathrm{i} \tilde{\omega} x}+ A^{+}_{h} e^{ \mathrm{i} \tilde{\omega} x},  &  x\rightarrow - \infty,
\\
e^{ \mathrm{i} \omega x},&   x\rightarrow + \infty.  \end{array}
\end{align}
If we express our general solution $\psi$ into this basis of solutions ($u_{h},u_{\infty}$), we can obtain the following relations between  the coefficients:
\begin{align}\label{coef1}
A^{+}_{h}&=\frac{\omega}{\tilde{\omega}}A^{-}_{\infty}, \\
\frac{A^\text{in}}{A^\text{wall}}&=\frac{\tilde{\omega}}{\omega}\left(A^{+}_{h}-A^{-}_{h}K e^{- 2 \mathrm{i} \tilde{\omega}x_0 }\right), \label{coef2} \\
A^\text{wall}&=\frac{\omega}{\tilde{\omega}}\frac{\left( A^{-}_{\infty}A^\text{out}-A^{+}_{\infty}A^\text{in} \right)}{K e^{- 2 \mathrm{i} \tilde{\omega}x_0  }} \label{coef3}.
\end{align}
These are the basic relations that are needed for the direct scattering problem calculation and further discussions about their derivation can be found in \cite{Torres:2022bto}.

For the direct problem, we first need to evaluate the set of coefficients $(A^{+}_{h},~ A^{-}_{h},~A^\text{+}_{\infty},~A^{-}_{\infty})$. 
This is done by numerically evolving the $u_{h}$ solution from the wall at $x_{0}$ to infinity, and vice versa, with the $u_{\infty}$ solution. 
With the evaluated coefficients, and by using Eqs.~\eqref{coef1},~\eqref{coef2} and~\eqref{coef3}, we can then determine the coefficients $(A^\text{wall},~ A^\text{in},~ A^\text{out})$, and, accordingly, the transmission and reflection coefficients $T$ and $R$, respectively.

Finally, the transmission and reflection coefficients, calculated here by means of the direct numerical problem, provide the starting point to study the inverse method that we consider within this framework. 
In principle these are the observables that could be obtained through future laboratory experiments of analog gravity systems.

\subsection{Inversion of Bohr-Sommerfeld rule and Gamow formula}\label{BS_Gamow}

With the numerical results for reflection and transmission coefficients available now, we will outline the different steps of the inverse problem method. 
We start with a high-level description of the main idea, and explain more specific details afterward.

The first step is to identify the location of resonance peaks as an approximation for the energies of the quasistationary states $E_{0n}$. 
Assuming that the Bohr-Sommerfeld rule eq.~\eqref{cBS} is a good approximation, it is known \cite{lieb2015studies,MR985100} that it can be used to reconstruct the ``width'' $L_1(E)$ of the potential well as a function of the energy via
\begin{align}\label{Excursion}
L_1(E) = x_1(E)-x_0(E) = \frac{\partial }{\partial E} I(E),
\end{align}
where $I(E)$ is the so-called inclusion and given by
\begin{align}\label{Inclusion}
I(E) = 2 \int_{E_\text{min}}^{E}\frac{n(E^\prime)+1/4}{\sqrt{E-E^\prime}} \mathrm{d}E^\prime.
\end{align}
Here $E_\text{min}$ is the minimum of the potential defined by extrapolating where $n(E)+1/4=0$. 
Note that the potential cannot be uniquely reconstructed, but instead there are infinitely many potentials with a given condition on their turning points. 

The second step is to combine the inversion of Gamow's formula for a two turning point potential barrier with the information about the potential well, which has been derived in Ref.~\cite{Volkel:2017kfj} for a single barrier next to a reflective boundary condition and in Ref.~\cite{Volkel:2018hwb} for quasistationary states trapped between two potential barriers. 
The inversion of the Gamow formula allows one to connect the transmission through a single potential barrier with the width of the barrier. 
This was first shown in \cite{1980AmJPh..48..432L,2006AmJPh..74..638G} and is given by
\begin{align}\label{widthbarrier}
L_2(E) &= x_1(E)-x_0(E) 
\\
&=  \frac{1}{\pi} \int_{E}^{E_\text{max}} \frac{\left(\text{d}T(E^\prime)/\text{d}E^\prime \right)}{T(E^\prime) \sqrt{E^\prime -E}} \text{d} E^\prime.
\end{align}
Note that $T(E)$ here is not the same as the full measured transmission coefficient, as the latter one includes the net result of the potential and the reflective wall combined together. 

To circumvent this limitation, we developed a numerical fitting procedure that provides an effective transmission through the potential barrier individually, which can then be used for the reconstruction of the barrier. 
This numerical procedure starts from the total transmission through the barrier and reflective wall and isolates the pure effect of the potential barrier. 
The final result obtained by this procedure is what one could use as the input for the inversion of the Gamow formula, given by Eq.~\eqref{widthbarrier} to reconstruct the barrier with the additional information coming from the potential well. 
However, in order to obtain robust results, one needs to slightly modify the transmission for energies close to the peak of the barrier, which we outline in the sequence. 
Finally, although we have not faced problems from possible low-energy inaccuracies from the WKB method, it could be a problem in other cases. 
In that case, it may be useful to extrapolate the low-energy transmission with analytic functions that do not cause so-called ``overhanging cliffs'' in the corresponding potentials; see Ref.~\cite{Volkel:2018czg} for a related study on such analytic extensions.

\subsubsection{Treatment at energies close to the barrier peak}

Because of the reduced validity of the Bohr-Sommerfeld rule and Gamow formula for energies around the peak of the barrier, we complement the close vicinity of the maximum of the potential barrier with a parabolic approximation
\begin{align}\label{parabollicfitting}
V_{\text{parabolic}}(x)=E_\mathrm{max}+a(x-x_\mathrm{max})^{2}.
\end{align}
Here the two relevant free parameters ($E_\mathrm{max}$ and $a$) are directly obtained from fitting the analytic form of the transmission to the numerical one; see Appendix of Ref.~\cite{Volkel:2019ahb}. 
With the estimate of $E_\mathrm{max}$, one can now compute $L_2(E)$ to obtain width-equivalent potentials $V_{L_2}(x)$. 
Finally, we define the effective reconstructed barrier to be a smooth interpolation between the two potentials
\begin{align}\label{transmissionfitting}
V_{\text{eff}}(x;x_\mathrm{int},\lambda)=&V_{\text{parabolic}}(x) \left( \frac{1}{2}-\frac{1}{2}\tanh[\lambda (x-x_\mathrm{int})]\right) \nonumber \\ +V_{L_2}(x)&\left(\frac{1}{2}+\frac{1}{2}\tanh[\lambda (x-x_\mathrm{int})] \right),
\end{align}
where $\lambda$ controls the ``smoothness'' of the transition between the two connected curves $V_{\text{parabolic}}(x)$ and $V_{L_2}(x)$ in $x_\mathrm{int}$, where these two curves intersect. 
Directly approximating the maximum of the potential with a parabola improves the reconstruction, because the inversion of the Gamow formula used to derive $L_2(E)$ is only valid for energies below the barrier peak. The choice of the matching point where the two curves shall intersect is done by optimizing the determination of the position of the parabola maximum in x axis, while looking for an intersection point where the functions to be matched have the same value and same slope.

\subsection{Analysis of the transmission}\label{analysis_transmission}

Knowing the transmission/reflection coefficients, we now outline how exactly we analyze it to provide the necessary input for the semiclassical inversion methods. 
These quasistationary states, as well as the transmission, are only related to the potential barrier and the reflective wall.

\subsubsection{Extracting quasistationary states from transmission}\label{methods_lorentzian}

Given the numerical transmission curve as the starting point of our analysis, we need to extract the spectrum of quasistationary modes. 
They are imprinted in the locations and widths of the resonance peaks. 
For example, the real part of the mode energies $(E_{0n})$ are the energy values at the local center of the peaks (or the local center of the small ``bumps'', in the low-reflectivity scenario), while the imaginary parts $(E_{1n})$ are related to the widths of those peaks. 

As we will discuss later, for some cases (high-reflectivity regime), the local center of the peaks are, with a very good approximation, the local maximum as well. For those cases, we extract the locations using a basic peak-finder algorithm, and then numerically fit a three parameter Lorentzian~\cite{riley_hobson_bence_2006} in a very close vicinity to it,
\begin{align}\label{eq_lorentzian}
f(E) =  \frac{T(E_\text{max, n}) \Gamma_n^2}{(E-E_\text{max, n})^2+\Gamma_n^2},
\end{align}
where $E_\text{max, n}$ is the location of the resonance peak and the real part of $E_n$, $\Gamma_n$ is the half width at half maximum, and $T(E_\text{max, n})$ is the peak value of the transmission at the resonant energy. 
The width $\Gamma_n$ of a certain peak will be related to the imaginary part/damping time of its respective mode. 
To increase the accuracy and optimize the algorithm, we iteratively refine the energy resolution in a more narrow region around a given peak. 

For the cases where the local maximums at the peaks are not a good approximation for the centers of the peak, a rather different approach is needed. 
This new approach is based on the analysis of the slope of the transmission and its variation within the peak. When passing through a peak/bump, the transmission's slope reaches a local maximum and it quickly decays into a local minimum (passing through zero, when there is a local maximum at this peak/bump). Accordingly, at the local center of the peak, the slope is decaying at the faster rate, so that the transmission's second-order derivative reaches a local minimum there. This way, we can estimate the local center of the small bumps by calculating the local minimum of their second derivative there.
We further discuss the low-reflectivity scenarios in Sec.~\ref{inverse_resultsDependencyonK}.

In all scenarios, we will be able to obtain the energies of the quasistationary modes $(E_{0n},E_{1n})$ for all different reflectivity regimes. 
These energies for the quasistationary states can then be solved for $n(E_0)$, interpolated, and then used as input for Eq.~\eqref{Inclusion}.

\subsubsection{Defining effective transmission through the barrier}\label{methods_effective_transmission}

As we demonstrate explicitly in Sec.~\ref{app_results}, one can use the transmission including the resonance peaks to construct an ``effective'' transmission that only captures the transmission through the barrier.  
To obtain this effective transmission, we first compute the envelopes connecting only the minima $T_\text{min}(E)$ and only the maxima $T_\text{max}(E)$ of the logarithm of the transmission curve and then construct an effective logarithmic transmission defined only by the envelopes $\log (T_\text{effective}) = (\log  (T_\text{max})(E)+\log (T_\text{min})(E))/2$. 
As can be seen in Fig.~\ref{envelope}, it is a very good approximation of the barrier transmission obtained in the case of perfect absorption at the core. 
Accordingly, this transmission will be the input for Eq.~\eqref{widthbarrier}.

\section{Application and results}\label{app_results}

In this section, we first outline the imperfect draining vortex system in Sec.~\ref{vortex}, and then show the results of our inverse method in Sec.~\ref{inverse_results}.

\subsection{Imperfect draining vortex model}\label{vortex}

In the following, we summarize the main details of the imperfect draining vortex as an analog of an extremely compact object. For more details, we
refer the reader to Ref.~\cite{Torres:2022bto}, where this model was outlined in more depth. 
The effective wave equation, which is the central piece of our analysis, can be written in the form
\begin{align}
\frac{\text{d}^2}{\text{d}x^2} \psi(x) - \bar{V}(r) \psi(x) = 0,
\end{align}
where the potential $\bar{V}(r)$ is given by
\begin{align}\label{potential}
\bar{V}(r) = - &\left(\omega - \frac{mC}{r^2} \right)^2 
\\
+ &\left(1-\frac{1}{r^2}\right) \left( \frac{m^2-1/4}{r^2} + \frac{5}{4r^4} \right),
\end{align}
where $x(r)$ is the so-called tortoise coordinate
\begin{align}\label{tortoise}
x(r) = r + \frac{1}{2} \log\left( \frac{r-1}{r+1}\right).
\end{align}
The rotational properties of the vortex are characterized by the constant $C$, and $m$ labels the harmonic decomposition used in the derivation of the effective wave equation for the radial part of the wave function. 
The latter one has a similar meaning as in the case of the Schwarzschild black hole. 
Note that, as in the case of rotating black holes, the potential becomes nontrivially $\omega$ dependent for rotating configurations (for $C\neq 0$). 
The absorption at the core of the vortex is modeled by defining a reflectivity $K$ through the interface surface, and it is defined by  the boundary conditions at $r = r_{h}(1+\epsilon)$, where $\epsilon$ is a very small number; we normalize our acoustic horizon radius with $r_{h}=1$. 

\begin{figure}
\includegraphics[width=1.0\linewidth]{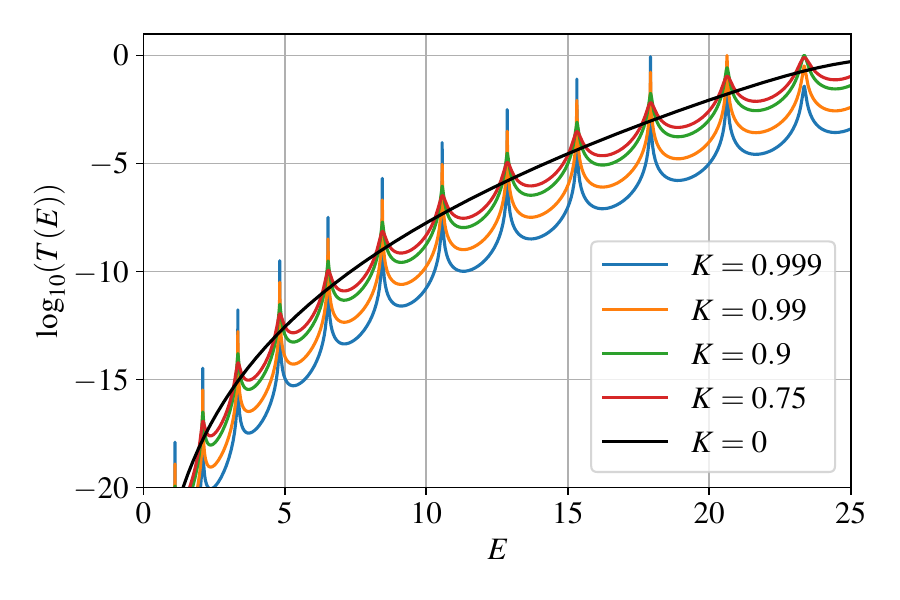}
\includegraphics[width=1.0\linewidth]{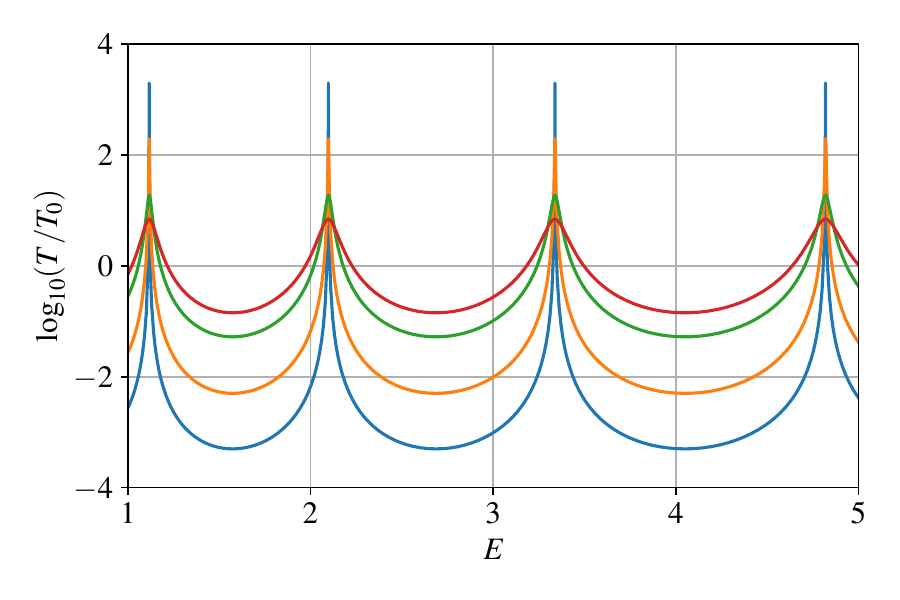}
\caption{Here we show different aspects of the transmission $T(E)$. Top: transmission for different values of $K=[0, 0.75, 0.9, 0.99, 0.999]$, $C=0$ and $m=10$. Bottom: same transmissions as before (same colors), but normalized with the one for no reflectivity $K=0$ denoted with $T_0$ and in a smaller $E$ range for better visibility of the resonance peaks. 
\label{fig_transmission_K}}
\end{figure}

\begin{figure}
\includegraphics[width=1.0\linewidth]{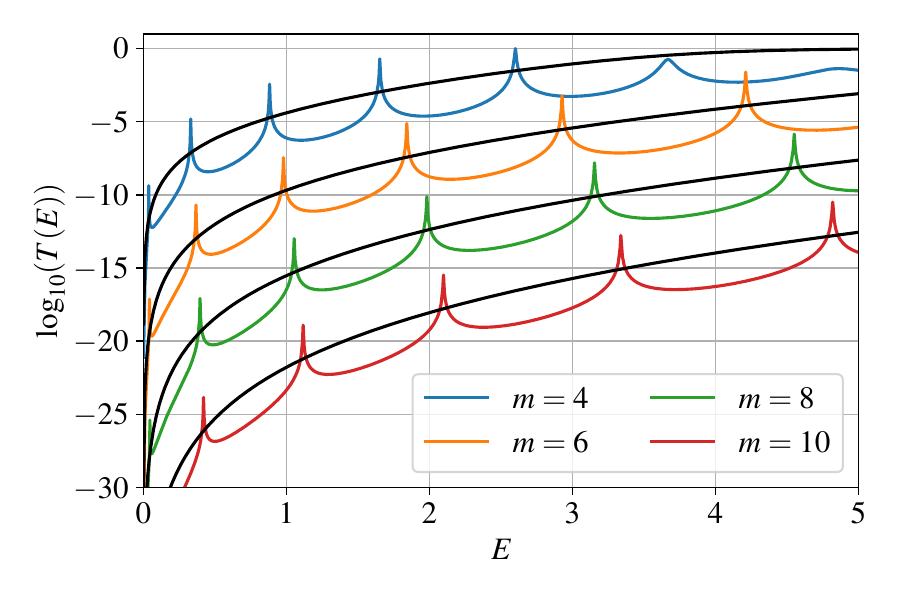}
\includegraphics[width=1.0\linewidth]{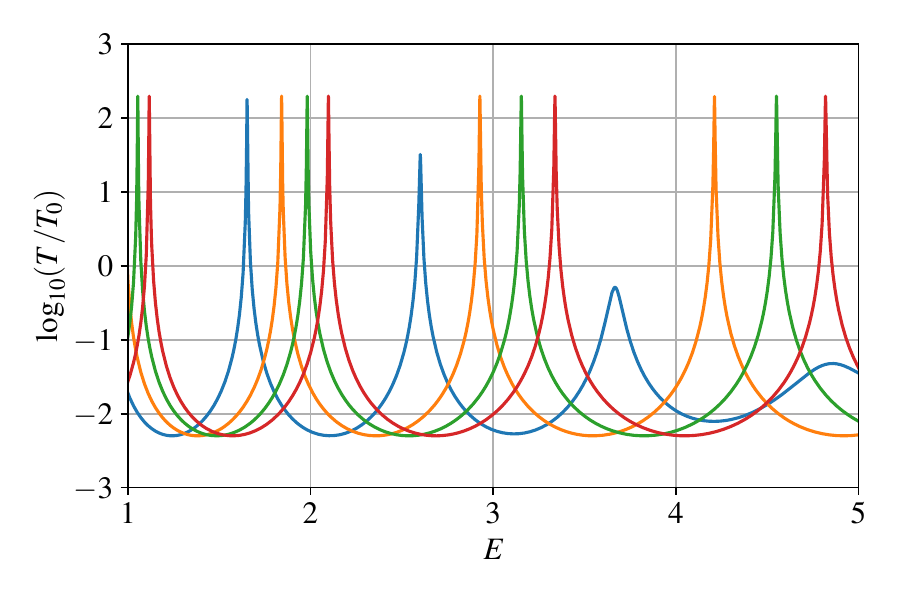}
\caption{Here we show different aspects of the transmission $T(E)$. Top panel: transmission for different values of $m=[4,6,8,10]$, $C=0$, and $K=0.99$. The $K=0$ case for each $m$ is shown for comparison (black lines). 
Bottom: same transmissions as before (same colors), but normalized with the ones for no reflectivity $K=0$ denoted with $T_0$ and in a smaller $E$ range for better visibility of the resonance peaks. 
\label{fig_transmission_m}}
\end{figure}

As applications for our inverse method, we have used the numerical setup described in Sec.~\ref{methods} to generate transmission curves as a function of energy. 
From now on, we assume that the reflective wall is defined as the cylindrical surface (since we are in a $2+1$-dimensional scenario) with a radial distance of the center defined by the value $(1+\epsilon)$, where $\epsilon=2e^{-20}$. 
According to Eq. \eqref{tortoise}, this implies that the tortoise coordinate is given by $x_{0}=-9$ at the reflective wall. 
In order to study different analog system realization, we choose several reflectivity values $K = [0, 0.75, 0.9, 0.99, 0.999]$, several harmonics $m=[4,6,8,10]$, and $C=0$. 
To make the impact of each parameter more clear and to avoid a plethora of various combinations, we vary only one of the parameters at a time and set the other ones to default values. 
The transmission curves for different reflectivity values are shown in Fig.~\ref{fig_transmission_K}. 
As one would expect, the resonance peaks become more dominant for $K \rightarrow 1$ and vanish in the limit $K \rightarrow 0$, but their location remains extremely similar. 
Varying the harmonic parameter $m$ yields transmission curves provided in Fig.~\ref{fig_transmission_m}. 
Note that $m$ changes the height of the potential barrier, which mainly controls the number of resonance peaks, but only mildly impacts their separation. 

\subsection{Reconstruction of potential and reflectivity}\label{inverse_results}

With the transmission curves of the previous section, we now apply the inverse methods introduced in Sec.~\ref{methods}. 
We first show and discuss our results for varying the harmonic number $m$ in Sec.~\ref{inverse_resultsDependencyonm} and then study the reflectivity $K$ in Sec.~\ref{inverse_resultsDependencyonK}.

\subsubsection{Dependency on harmonic $m$}\label{inverse_resultsDependencyonm}

\begin{figure}
\includegraphics[width=1.0\linewidth]{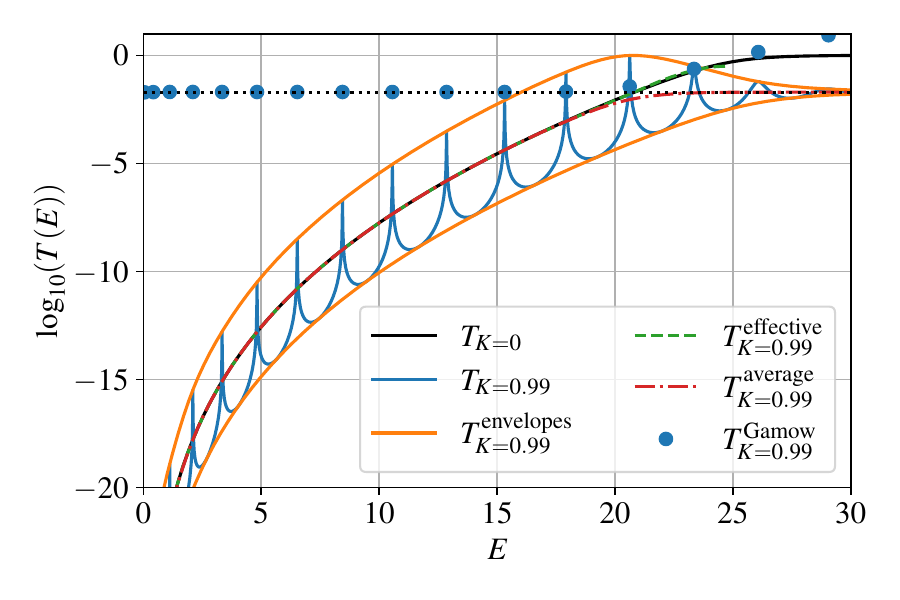}
\includegraphics[width=1.0\linewidth]{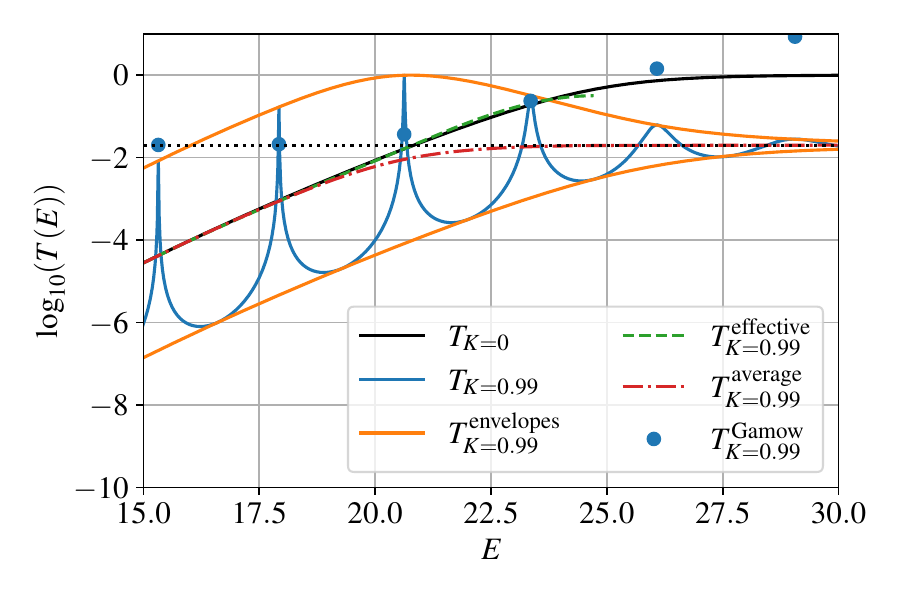}
\caption{Here we compare the transmission $T_{K=0.99}(E)$ for $m=10$ and $C=0$ (blue solid) with ${T}_{K=0}(E)$ (black solid line). 
The corresponding envelopes ${T}^\mathrm{envelopes}_{K=0.99}(E)$ (orange solid line) and the average transmission ${T}^\mathrm{average}_{K=0.99}(E)$ (red dash-dotted line) defined by the two envelopes that are shown as well. 
It is evident that the average transmission is a very accurate approximation for ${T}_{K=0}(E)$ until $E\approx 20$, where it plateaus toward around $10^{-2}$ (black dotted line), which corresponds to $1-K^2$. 
Here, the effective transmission ${T}^\mathrm{effective}_{K=0.99}(E)$ (green dashed line) follows $T_{K=0}(E)$ closely until around the maximum of the potential barrier (around $E=25$). 
$T^\mathrm{Gamow}(E)$ (blue points) are the transmissions obtained from the resonance peaks and Gamow formula, see main text. 
Bottom: we show the same system as in the top, but for a smaller energy range for better visibility of details. 
\label{envelope}
}
\end{figure}

\begin{figure}
\includegraphics[width=1.0\linewidth]{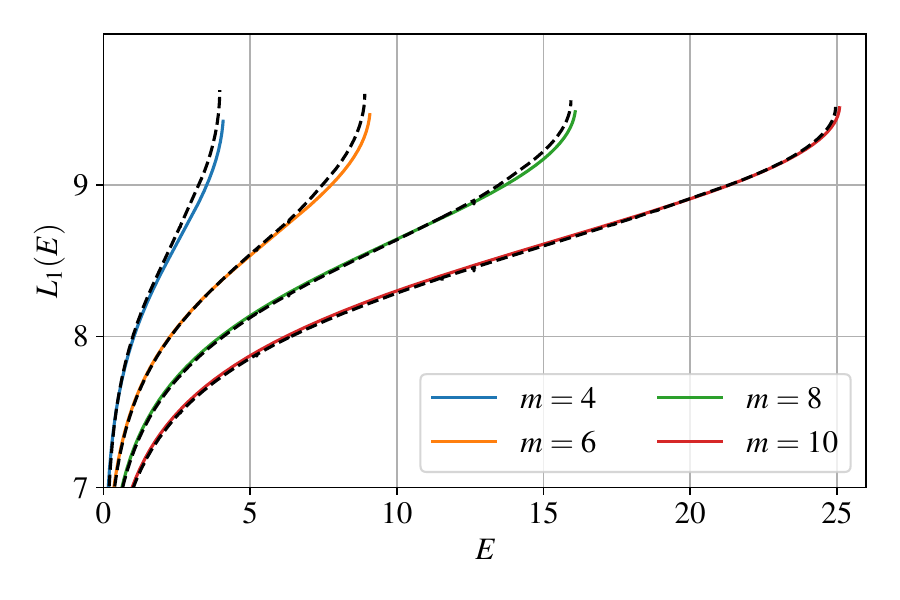}
\includegraphics[width=1.0\linewidth]{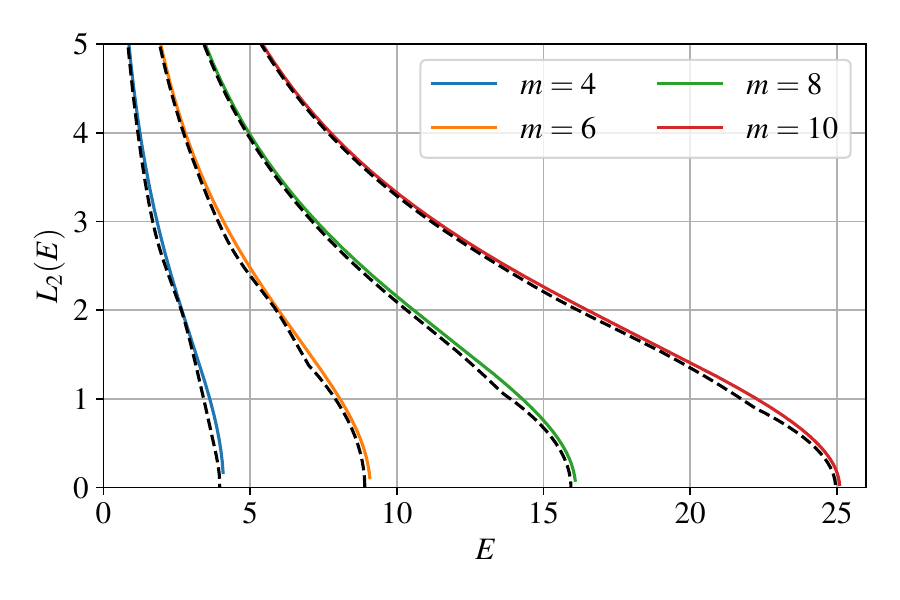}
\caption{In this plot we show the exact (colored solid) and reconstructed (black dashed) widths $L_1(E)$ (top panel) and $L_2(E)$ (bottom panel) for $K=0.99$ and $m = [4,6,8,10]$. \label{results_L1L2_m}}
\end{figure}

In the following, we demonstrate the various steps that have been explained in Sec.~\ref{methods}. 
We start our analysis with the transmission curves from Fig.~\ref{fig_transmission_m} (for $K=0.99$ and $m = [4,6,8,10]$). 
To obtain an accurate estimate for the location and widths of the resonance peaks, we could first normalize it with the $K=0$ transmission. 
If the $K=0$ transmission is not available, e.g., because such a case could not be realized in an experiment, it can also be approximated with high accuracy from constructing $T_\mathrm{effective}(E)$ from envelopes, as discussed in Sec.~\ref{methods_effective_transmission}. 
In the following, we assume the latter is the case and do not make explicit use of any $K=0$ knowledge. 
The result of the envelope construction is shown in Fig.~\ref{envelope}, which clearly demonstrates the excellent agreement between the effective transmission and the $K=0$ transmission, at least until it reaches energies close to the maximum of the potential ($E \approx 20\sim 25$). 
Using the location of the resonance peaks $E_{0n}$, we can invert the relation for $n(E_{0n})$, interpolate it, and use it as input for Eq.~\eqref{Inclusion}, which then enters Eq.~\eqref{Excursion}. 
This yields the width of the cavity $L_1(E)$, which we report in the bottom panel of Fig.~\ref{results_L1L2_m}, 
and concludes the reconstruction of the cavity properties. 

Next, we use the effective transmission $T_\mathrm{effective}(E)$ to compute the width of the barrier $L_2(E)$ via Eq.~\eqref{widthbarrier}. 
Because $T_\mathrm{effective}(E)$ deviates from the $K=0$ transmission close to the potential maximum (depending on the value of $K$), we use the Gamow formula Eq.~\eqref{E1n_1} and the width of the resonance peaks $\Gamma_n = E_{1n}$ to define $T_2(E)$. 
The Gamow formula relates those values with the sum of transmission $T_1(E) + T_2(E)$. 
Note the presence of the integral over the potential well, which can only be computed using our reconstructed width $L_1(E)$. 
Because of the nonuniqueness of the reconstructed potentials from $L_1(E)$, we can construct any potential with such a turning point relation to carry out the integration numerically. Since the transmission is constant through the wall $T_1(E)=1-K^2$, and the transmission through the barrier  $T_2(E)$ is exponentially smaller for lower energies, we can assert that, in the low-energy regime, the sum $T_1(E) + T_2(E)$ tends to $T_1(E) + T_2(E)\approx T_1(E)=1-K^2$. This helps us infer the transmission through the wall $T_1(E)$ and its associated reflectivity $K$. With these values, we can infer the behavior of $T_2(E)$ for higher energies if we use $T_2(E)= T_1(E) + T_2(E)-(1-K^2)$, where $T_1(E) + T_2(E)$ is obtained by Gamow formula [Eq.~\eqref{E1n_1}]. This procedure gives us some points slightly below the blue dots shown in Fig.~\ref{envelope}. If we interpolate those points, we obtain the green dashed line, which can be used to properly continue the $T_\mathrm{effective}(E)$ in the energy domain where the envelopes' mean failed to approximate the transmission through the barrier ${T}_{K=0}(E)$. 
 
Smoothing $T_2(E)$ with $T_\mathrm{effective}(E)$ we capture the barrier transmission for low and maximum energies and, finally, use it in Eq.~\eqref{widthbarrier} to obtain $L_2(E)$, which we report in the bottom panel of Fig.~\ref{envelope}. Because the $L_2(E)$ integration requires the knowledge of $E_\mathrm{max}$, which is not known \textit{a priori}, we used the parabolic transmission \eqref{parabollicfitting} approximation to fit $T_2(E)$ transmission in a range that can initially be estimated from where the transmission starts to plateau. 

\begin{figure}
\includegraphics[width=1.0\linewidth]{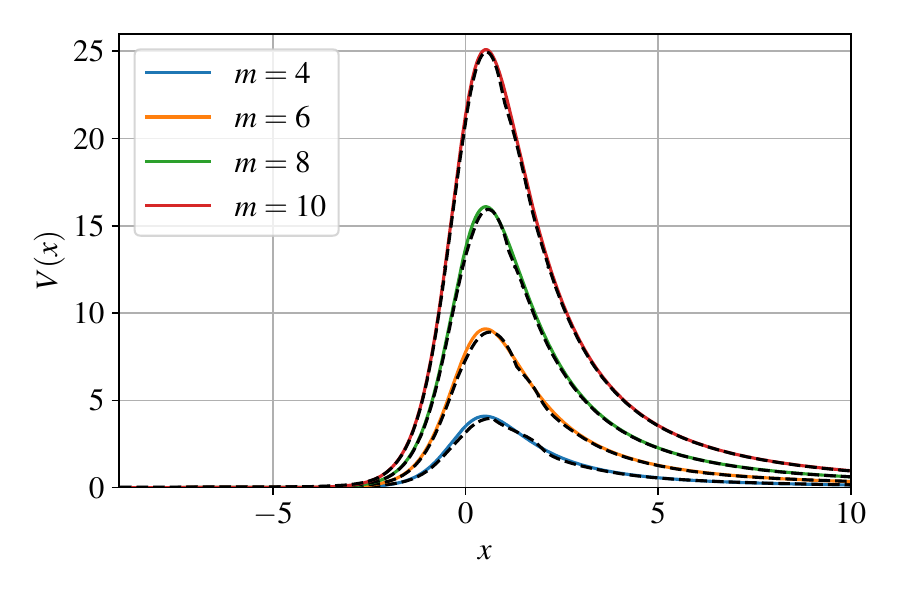}
\caption{Here we show the true (colored lines) and reconstructed (black dashed) effective potentials $V(x)$ for $K=0.99$ and $m = [4,6,8,10]$. 
The location of the effective core's surface is at $x=-9$. 
\label{results_V_m}}
\end{figure}

With two relations $L_1(E), L_2(E)$ for three turning points $x_0, x_1, x_2$, one needs to provide a third relation to define a specific potential. 
The natural choice in our problem is to assume that the location of the reflective wall does not depend on the energy, and thus, we set $x_0$ to be some constant. 
The only freedom in choosing the constant is a coordinate shift, which is not relevant for the underlying properties of the system. 
Finally, we report the reconstructed potentials defined by this choice in Fig.~\ref{results_V_m}. 
As can be seen in both figures, the overall accuracy of the reconstruction improves with larger values of $m$. 
Because $m$ mostly controls the height of the barrier, and thus the number of quasistationary states that appear as resonance peaks, one should expect the reconstruction to be more accurate because more information can be used for the interpolation of the spectrum and effective transmission. 
Furthermore, the underlying WKB-based methods are expected to be most accurate for the quasistationary states that are located not too close to the minimum of the potential, and not too close to the maximum of the barrier. 

\subsubsection{Dependency on reflectivity $K$}\label{inverse_resultsDependencyonK}

What remains is the reconstruction of the corresponding reflectivity parameters $K$. 
We assume that the wall's reflectivity $K$ is energy independent and thus the same for all different incident wave frequencies. 
In this case, the transmission through the wall is also a constant and given by
\begin{align}\label{transwall}
T_1(E)=1-K^2.
\end{align}
Accordingly, as an outcome of applying the Gamow formula Eq.~\eqref{E1n_1}, one can obtain the sum of the value above, for the transmission through the wall, with the transmission $T_2(E)$ through the potential barrier as if the wall would be perfectly absorbing. 
The sum $T_1(E)+T_2(E)$ is dominated by $T_1(E)$ for low energies, because $T_2(E)$ becomes exponentially small. 
This fact can be illustrated graphically in Fig.~\ref{envelope}. 
If we look at the blue dots [the Gamow points $T_1(E)+T_2(E)$], we can see that they start to plateau as we decrease the energy. 
This plateau gives us the constant value of $T_1(E)=1-K^2$. 

\begin{figure}
\includegraphics[width=1.0\linewidth]{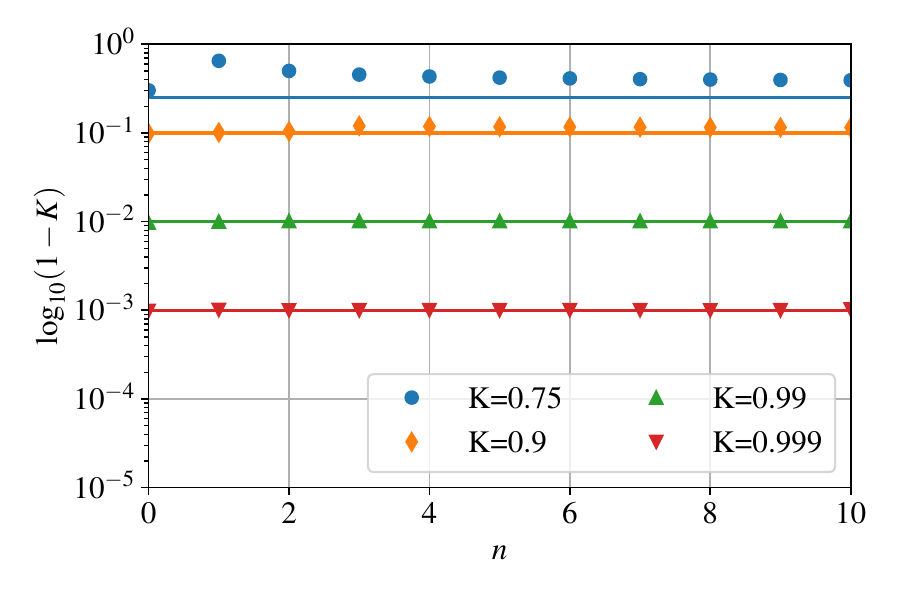}
\caption{Reconstruction of the different reflectivity parameters $K=[0.75, 0.9, 0.99, 0.999]$ with the inverse method using the resonance peaks and Gamow formula. 
The solid lines correspond to the exact values, while the different points are the reconstructed values using the $n$ resonance peak. 
\label{results_K}}
\end{figure}

We show the reconstructed values of $K$ in Fig.~\ref{results_K}. 
Here, the $x$ axis labels the $n-\mathrm{th}$ quasistationary state that has been used in the Gamow formula Eq.~\eqref{E1n_1}. 
One can observe that for values of $K$ close to 1, the reconstruction is very accurate. 
In this case the transmission through the wall is much smaller than the one through the barrier and the resonance peaks can be very accurately extracted. 
For smaller values of $K$, the reconstruction looses accuracy, and deviates from the correct injection by $10\,\%\sim20\,\%$ for $K=0.75$. 
To investigate this, we tried several improvements. 
First, even when the $T_2(E)$ contribution is included in the Gamow formula (by using the effective transmission extrapolated), the results for $K=0.75$ do not change significantly, especially not for moderate values of $n$, where the approximation is excellent. 
Second, we also checked whether fitting all resonance peaks simultaneously can give better results, because peaks start to overlap and results may get biased. 
However, also in this case we do not report improvements, as we fit the resonance peaks in a close vicinity around the maximum, where the impact of the other peaks is mainly absorbed by the value of the transmission at each maximum, $T(E_\mathrm{max})$ Eq.~\eqref{eq_lorentzian}, and does not impact $\Gamma_n$ significantly. 
Finally, the alternative and more direct approach to determine $K$ is from $T(E)$ via Eq.~\eqref{transwall} evaluated for energies beyond the maximum of the barrier, since then the effect of the barrier becomes negligible. 
We note that the latter approach is complementary to using the widths of the resonance peaks. 
Which of the two approaches yields more accurate results when applied to real data with measurement uncertainties remains for future work. 

\section{Conclusions}\label{conclusions}

Analog gravity systems may provide unique and controlled measurements of their key properties, which are not accessible from their astrophysical counterparts. 
The novel method that we developed in this work is based on the extension of semiclassical methods and is tailored to study measured transmission/reflection coefficients of analog exotic compact objects. 
The outcome of the method is the reconstruction of the effective potential, which captures the dynamical properties of the system, as well as the reflectivity coefficient describing the internal boundary condition. 
In this work, we chose the imperfect draining vortex model suggested in Ref.~\cite{Torres:2022bto} as one example. 
First, we obtained the transmission/reflection coefficients with accurate (numerical) methods to explore different properties of the system, in particular, the impact of the reflectivity $K$ and different angular numbers $m$. 
These results were then used as ideal measurements of a future experiment to demonstrate the capabilities of the new (semiclassical) method. 

Our main findings are as follows. 
The reconstruction of the effective potential becomes very accurate with increasing angular numbers $m$, which is expected from the validity of the underlying WKB theory. 
This is also related to the fact that for the same location of the core, increasing $m$ yields more resonance states, and thus more information used for the interpolation of the inclusion ~\eqref{Inclusion} that is needed to reconstruct the width of cavity. 
The reconstruction of the reflectivity coefficient $K$ through the width of the resonance peaks becomes more accurate when it approaches unity, which corresponds to the full reflection case. 
This may also be expected, because large values of the reflectivity result in more prominent resonance peaks. 
Although inverse problems are often not uniquely solvable (typically not limited by the chosen methods), we suggested physically motivated assumptions that allow one to reconstruct the effective potential. 
We want to stress that because the input of our method has been computed with accurate numerical methods, but the reconstruction is based on semianalytic results, comparing the original potential with its reconstruction is not circular, but indeed self-consistent. 
Because of the explicit energy dependence of the potential for rotating configurations (for $C\neq0$), we have focused on $C=0$ and leave the conceptually more involved inverse problem of the energy-dependent potential for a separate work. 

Since our method is based on modifying similar approaches for the inverse problem of quasistationary states~\cite{Volkel:2017kfj,Volkel:2018hwb} and Hawking radiation~\cite{Volkel:2019ahb}, we also want to briefly compare some aspects. 
Although the knowledge of the transmission/reflection coefficients does not rely on the knowledge of the spectrum of quasistationary states, our method partially relies on identifying them indirectly.  
Thus, for specific model parameters that only provide very few such states, our method is not very accurate. 
However, since increasing angular numbers yields larger potential barriers, they also yield potentials with more quasistationary states. 
This means, if one is experimentally able to measure transmission/reflection coefficients of large enough angular numbers, our method can, even in such cases, always be used. 
In the context of astrophysical objects, this is not easily possible, as standard binary mergers mostly excite small angular numbers, which undermines the opportunities of studying analog gravity systems. 

We conclude with a comment on measurement uncertainties. 
Throughout this work we assumed that the transmission/reflection coefficients can be provided with pristine accuracy. 
However, any real experiment will come with statistical and systematic uncertainties, which may need to be taken into account. 
This could, for example, be done by repeating the reconstruction procedure for different realizations of the transmission coefficient that represent the statistical uncertainties of the measurements. 
At the same time, these uncertainties may not be relevant for all energy ranges, since the transmission varies over many orders of magnitude. 
We leave a detailed study of these aspects for future work. 

\acknowledgments
S.~A. acknowledges funding from Conselho Nacional de Desenvolvimento Cient\'ifico e Tecnol\'ogico (CNPQ)-Brazil and Coordena\c{c}\~ao de Aperfei\c{c}oamento de Pessoal de N\'ivel Superior (CAPES)-Brazil. 
S.~H.~V. acknowledges funding from the Deutsche Forschungsgemeinschaft (DFG): Project No. 386119226. V.~B.~B.s partially supported by the Conselho Nacional de Desenvolvimento Científico e Tecnológico (CNPq)-Brazil, through the Research Project No. 307211/2020-7.

\bibliography{literature}

\end{document}